\documentstyle[12pt]{ioplppt}
 
\begin{document}
\input amssym.def
\input amssym
\jl{6}

\title{Fuzzy spacetime from a null-surface version of GR}

\author{
	S Frittelli\dag\ftnote{3}{e-mail: simo@artemis.phyast.pitt.edu},
	C N Kozameh\ddag\ftnote{4}{e-mail: kozameh@fis.uncor.edu}     ,
	E T Newman\dag\ftnote{5}{e-mail: newman@vms.cis.pitt.edu}      , 
	C Rovelli\dag\ftnote{6}{e-mail: rovelli@pitt.edu}       and 
	R S Tate\dag\ftnote{7}{e-mail: rstate@minerva.phyast.pitt.edu}
	}
\address{
	 \dag Department of Physics and Astronomy, 
		University of Pittsburgh, 
		Pittsburgh, PA 15260				      }
\address{
	 \ddag FAMAF, Universidad Nacional de C\'{o}rdoba, 
		5000 C\'{o}rdoba, 
		Argentina	
        }
\date{\today}
\maketitle

 	{\it Dedicated to Andrzej Trautman, who so early on saw the
	importance of characteristic surfaces in general relativity.}

\begin{abstract}
	The null-surface formulation of general relativity -- recently
	introduced -- provides novel tools for describing the
	gravitational field, as well as a fresh physical way of viewing
	it. The new formulation provides ``local'' observables
	corresponding to the coordinates of points --- which constitute
	the spacetime manifold --- in a {\em geometrically defined
	chart\/}, as well as non-local observables corresponding to
	lightcone cuts and lightcones.  In the quantum theory, the
	spacetime point observables become operators and the spacetime
	manifold itself becomes ``quantized'', or ``fuzzy''.  This
	novel view may shed light on some of the interpretational
	problems of a quantum theory of gravity.  Indeed, as we discuss
	briefly, the null-surface formulation of general relativity
	provides (local) geometrical quantities --- the spacetime point
	observables --- which are candidates for the long-sought
	physical operators of the quantum theory.
\end{abstract}


\section{Introduction} 

The issue of the relationship or unification of the general theory of
relativity with quantum theory was raised early on, and remains a
subject of active research~\cite{chris}. Two basic types of problems
have been at the core of the work: 1. In a purely formal manner, can
(or how can) a quantum theory of general relativity (GR) be constructed
in a mathematically consistent fashion? 2. If, in fact, a theory can be
so constructed, can (or how can) it be given a reasonable physical
interpretation, i.e., what relationship does it have to physical
reality? In particular, what possible physical meaning could be given
to a ``quantum spacetime"?

According to some these problems can be solved by simply applying the
ideas of quantum theory to GR or to one of its
extensions~\cite{some,canonical}.  There are others who feel that both
the mathematical and conceptual problems are so overwhelming that new
physical ideas (perhaps radical) must be introduced to unify GR with
quantum theory - that there must be a true unification or amalgamation
of the ideas of GR with the ideas of quantum theory, rather than the
``simple" subjugation of GR to the machinery of quantum
theory~\cite{others}.

Without strongly taking sides in the debate (the authors among
themselves have differences of opinion) we want to point out that
by viewing the classical Einstein equations in an unorthodox manner and
following a ``natural" trail towards the ``quantization" of this
unorthodox version of GR, we appear to have been led to a new point of
view towards the subject. The new view essentially states that the
spacetime points themselves must be ``quantized", i.e., turned into
operators with commutation relations, etc. It is not the fields on a
manifold that must be quantized (metric, etc.), but is (in some sense)
the manifold idea itself that must be changed. This is not simply an
empty conjecture; there is in principle (in full theory) and in
practice (in linear theory) a means to calculate the commutators
between individual spacetime points. We are not claiming anything
profound nor are we even advocating an approach to the two major
questions posed above.  We, however, feel that the new point of view
has the potential for shedding light on the problem of the ``meaning of
quantum gravity" and thereby being possibly of importance.

In Sec.~\ref{null-surface theory} we give a brief introduction to a new
(classical) point of view to GR, referred to as a null-surface version
of GR (where the basic variable has been shifted from the metric,
connection, or other -- field --, to families of characteristic
surfaces) while, in Secs.~\ref{vacuumaxwell} and \ref{lingr}, we
discuss what happens when we try to turn the new point of view into a
quantum theory.  In Sec.~\ref{conclusion} we discuss possible meanings
and ramifications of these ideas.

This is a report of work very much in progress and as such we will
mainly ``wave our hands", describe what we have in mind and give a
minimum number of equations and no proofs.


\section{Null-Surface View of GR}	     \label{null-surface theory}

During the past few years a new formulation of classical general
relativity (specially suited for asymptotically flat spacetimes) has
become available~\cite{nullsurface}. In this formulation, the emphasis
has shifted from the usual type of field variable (metric, connection,
holonomy, curvature, etc.)  to, instead, families of three-dimensional
surfaces. On a manifold with no other structure, equations are given
for the determination of these surfaces.  From the surfaces themselves,
by differentiation, a (conformal) metric can be obtained; the surfaces
themselves are then automatically characteristic surfaces of this
conformal metric. The equations simultaneously determine a choice of
conformal factor such that the metric automatically satisfies the
vacuum Einstein equations. In other words, the vacuum Einstein
equations are formulated as equations for families of surfaces and a
single scalar conformal factor. In our present discussion we will be
primarily interested in only the characteristic surfaces, though, in
the actual formalism, the conformal factor plays an essential role in
the determination of the surfaces. Our point of view will be that we
have already integrated the equations for both -- the surfaces and the
conformal factor -- and we are interested now in information that we can
retrieve from knowledge of the surfaces.

Consider asymptotically flat spacetimes with (future) null infinity
${\cal I}^+=S^2\times {\rm I\!R}$. Let $\zeta$ be a complex
stereographic coordinate\footnote{
	In this paper, when we write a function of complex variables,
	we make no assumption of it being holomorphic.  Consequently,
	functions that are of the form $f(\zeta,\bar\zeta)$ will be
	written simply as $f(\zeta)$. Consistently, we will write 
	$\zeta$ instead of $(\zeta,\bar\zeta)$, except when confusion
	could arise.
	}
on $S^2$ which labels the generators of ${\cal I}^+$, and let $u\in
{\rm I\!R}$ be an appropriately normalized parameter along the
generators. Thus $(u, \zeta)$ are the Bondi coordinates on ${\cal
I}^+$. The free data for Einstein's equations is then associated with a
connection (the Bondi shear) on ${\cal I}^+$, and can be specified by
the choice of a complex spin-weight-2 field $\sigma(u,\zeta)$ on ${\cal
I}^+$.  The space of all such fields $(\sigma)$, together with the
appropriate symplectic structure, constitutes the reduced phase space
of general relativity~\cite{aa:aq}.

The characteristic surfaces of the spacetime are described as follows:
We obtain a function (our fundamental variable)
$Z(x^a,\zeta;\,[\sigma])$ as a solution of certain differential
equations involving the given data $[\sigma]$. (The square brackets
$[\;]$ denote functional dependence.) The $x^a$ appearing in the
argument of $Z$ are local coordinates on the spacetime manifold in an
arbitrary chart. They appear as (integration) parameters in the
solutions of the differential equations that determine $Z$.  For fixed
values of $(u,\zeta)$, the equation
  \begin{equation}					\label{lightcone}
 			Z(x^a,\zeta;\,[\sigma]) = u
  \end{equation} 
describes a characteristic (or null) surface in terms of the given
local chart, $x^a$, on our manifold. In fact, the null surface is the
past lightcone of the point $(u,\zeta)$ on ${\cal I}^+$. As the value
of $u$ varies (for fixed $\zeta$) we have a one-parameter foliation
(of a local region) by the characteristic surfaces. The $\zeta$ then
labels a sphere's worth of these null foliations; equivalently, for each
point $x^a$, as the $\zeta$ varies, we obtain a sphere's worth of
characteristic surfaces through $x^a$. An alternate interpretation of
the function $u=Z(x^a,\zeta;\,[\sigma])$ is that, for fixed point
$x^a$, it describes the lightcone cut of $x^a$; i.e, the intersection
of the future lightcone of $x^a$ with ${\cal I}^+$.

Assuming that the $Z$ satisfies our differential equations, one can
then, in a prescribed fashion, express a conformal Einstein metric in
terms of derivatives of $Z$~\cite{nullsurface}. For simplicity, a
special member of the conformal class can be chosen (in a natural\footnote{
	A special member of the conformal class is chosen by requiring
	that the coordinate $R$ introduced in (\ref{newcoordsr})
	becomes an affine parameter with respect to that representative
	of the conformal class~\cite{nullsurface}. 
	 }
fashion), yielding an explicit metric in terms of $Z$. This natural
representative of the conformal class {\it does not\/} satisfy the
Einstein equations. Nevertheless, once the conformal representative is
fixed, the conformal factor needed to transform it to an Einstein
metric can be determined as well.  In what follows, we assume that the
function $Z$ is always implicitly associated with an appropriate
conformal factor that guarantees an Einstein metric.  Notice, first,
that all the conformal information about the spacetime is contained in
the knowledge of $Z(x^a,\zeta;\,[\sigma])$ and, second, that the
(Einstein) conformal factor itself depends on the data.  (For the sake
of simplicity of presentation, we have slightly simplified the
discussion.  See \cite{nullsurface} for the details.)

Since, for each fixed value of $\zeta$, as $u$ varies, the $Z$
describes a foliation by null surfaces, we can associate a null
coordinate system with each such foliation. Though it is not at all
obvious (the proof is given in~\cite{nullsurface}) a characteristic
coordinate system is easily explicitly obtained from the $Z$ --- by
taking several $(\zeta,\bar\zeta)$ derivatives --- in the following
fashion:
 \numparts				             \label{allnewcoords}
  \begin{eqnarray}
	u    =              Z(x^a,\zeta,\,[\sigma]),  \label{newcoordsu}\\
     \omega  =         \eth Z(x^a,\zeta,\,[\sigma])   \label{newcoordso}\\
 \bar\omega  =     \bar\eth Z(x^a,\zeta,\,[\sigma]),  \label{newcoordsobar}\\
	 R   = \eth\bar\eth Z(x^a,\zeta,\,[\sigma]),  \label{newcoordsr}
  \end{eqnarray} 
 \endnumparts
where $\eth$ and $\bar\eth$ are (essentially) the $\zeta$ and
$\bar\zeta$ derivatives respectively~\cite{np66}. The geometrical
meaning of the coordinates $(u,\omega,\bar\omega,R)$ is extensively
discussed below.  Note that we automatically have a sphere's worth of
these characteristic coordinate systems (a single coordinate system for
every value of $\zeta$).

With the notation
  \begin{equation} 
 		   \theta^i	 = 
	     (
		   \theta^0 ,
		   \theta^+ ,
		   \theta^- , 
		   \theta^1 	    
	      ) 		 = 
	     (	          u , 
		   \omega   ,
	       \bar\omega   , 
			  R
	      )
  \end{equation} 
we have that 
  \begin{equation}				 
 	           \theta^i 			   = 
		   \theta^i  (x^a,\zeta;\,[\sigma]) 
							\label{newcoords} 
  \end{equation} 
is a coordinate transformation from the ``old'' coordinates $x^a$ to a
sphere's worth of null coordinate systems $\theta^i$. It should however
be stressed that (\ref{newcoords}) is much more than a set of
coordinate transformations. The $\theta^i$ contain the {\em full
information\/} about the solutions of the conformal Einstein equations,
through their dependence on the data $\sigma$.

Eq.\ (\ref{newcoords}) can be algebraically inverted to
express the local coordinates $x^a$ in terms of the $\theta^i$ and
$\zeta$
  \begin{equation}				\label{oldcoords} 
     x^a				  =
     x^a(   \theta^i, \zeta,\,[\sigma])	  \equiv 
     x^a(u, \omega,R, \zeta;\,[\sigma]).  
  \end{equation}
The inversion shows that, expressed in the chart $\theta^i$, the $x^a$
depend on the data.  Note that, like (\ref{newcoords}),
(\ref{oldcoords}) also contains the full information about the
solutions of the conformal Einstein equations; i.e., from
(\ref{oldcoords}) a metric conformal to an Einstein metric can be
obtained. Though this feature is basic, (\ref{oldcoords}) encodes other
information which is, at the moment, of more direct interest to us.
Before we proceed to the quantum theory, it is worthwhile to discuss
the meaning and the dual role of equations (\ref{allnewcoords}) and
(\ref{oldcoords}). We present three related and complementary
interpretations, which focus on three different sets of geometric
entities. For the purposes of the following discussion, we fix the data
$\sigma$ to some arbitrary value, thus fixing an Einstein 4-metric.

\subsection{Lightcone cuts}		\label{lcc}

Consider first Eq.~(\ref{lightcone}). We assume that $Z$ is known, as a
solution of our equations (equivalent to the Einstein equations).
Everything we say follows from the $Z$.

On the one hand, Eq.~(\ref{lightcone}) has been introduced as
describing null surfaces of the interior spacetime in terms of the
``old'' coordinates $x^a$, since, viewed as a function of the $x^a$,
(\ref{lightcone}) defines the past lightcone of the point $(u,\zeta)$
on ${\cal I}^+$: spacetime points $x^a$ which solve $Z(x^a,\zeta)=u$
lie on this lightcone.

However, there is another meaning (mentioned earlier), which extends to
the other $\theta^i$. We can view $Z(x^a,\zeta)$, for fixed $x^a$, as
describing a cut on ${\cal I}^+$ (the lightcone cut of $x^a$); i.e.,
$u$ as a function of $\zeta$. Consider a fixed $x^a$. For every fixed
value of $\zeta$ there is (in general) a single null geodesic
emanating from $x^a$ that reaches ${\cal I}^+$. This null geodesic can be
characterized by geometric information given on ${\cal I}^+$; three real
parameters labeling the point of intersection $(u,\zeta)$ and two
parameters specifying a direction from ${\cal I}^+$ (complex $\omega$,
geometrically a 2-blade). These parameters correspond to the five real
parameters $(u,\omega,\zeta)$ given by Eqs.\ (\ref{allnewcoords}). The
parameter $R$ (Eq.~(\ref{newcoordsr}) parametrizes the geodesic
labeled by $(u,\omega,\zeta)$ and specifies the interior points that
lie on it.  Eqs.\ (\ref{allnewcoords}) fix a one-to-one correspondence
between interior points and geometric structures on ${\cal I}^+$.  For the
moment, this is the viewpoint we will take for the meaning of
(\ref{allnewcoords}):
\begin{quotation}

   {\em For a fixed spacetime point $x^a$, (\ref{newcoordsu}) yields
	its lightcone cut on ${\cal I}^+$. (\ref{newcoordso}) yields the
	angle of intersection with the generator $\zeta$ of ${\cal I}^+$ of
	the null geodesic between $x^a$ and the generator $\zeta$, and
	(\ref{newcoordsr}) yields the corresponding non-affine geodesic
	distance from $x^a$ to ${\cal I}^+$. 
   }
\end{quotation}

(Note that $R$ also has a geometric meaning on ${\cal I}^+$; it is the
curvature of the lightcone cut at $\zeta$.)

\subsection{Spacetime points}		\label{spacetime points}

Alternatively to (\ref{newcoords}), we can think of its inverse,
namely (\ref{oldcoords}), as locating interior spacetime points by
following null geodesics inward, from ${\cal I}^+$, as follows. The
five-dimensional space of null geodesics (of the spacetime specified by
$\sigma$) is coordinatized by $(u, \omega, \zeta)$:  $(u,\zeta)$ label
the point on ${\cal I}^+$ and $\omega$ labels the generator of the past
lightcone; whereas $R$ parametrizes the individual geodesics. For fixed
values of $(u,\omega,R,\zeta)$, (\ref{oldcoords}) yields an interior
spacetime point.

\begin{quotation}

   {\em Given a fixed ``observation point'' $(u,\zeta) \in {\cal I}^+$,
	(\ref{oldcoords}) gives the coordinates (in the local chart) of
	the point which lies a parameter distance $R$ along the
	generator $\omega$ of the past lightcone of $(u,\zeta)$.
    }
\end{quotation}

Observe that a particular point $x^a$ can be reached by coming in along
many different null geodesics; i.e., there are many different
sets\footnote{Exactly $S^2$ worth, see (\ref{lccut}).} of parameters
$(u, \omega, \zeta)$ that will focus the null geodesics onto the same
point $x^a$.  {\em Spacetime points are thus identified by the set of
null geodesics that focus onto them\/}.

As an aside we also point out that (\ref{oldcoords}) written as
$x^a=x^a(R,u,\omega,\zeta,[\sigma])$ for arbitrary but fixed
$(u,\omega,\zeta)$, and variable $R$, is the parametric form for {\em
all\/} null geodesics.

\subsection{Lightcones}			\label{lightcones}

Finally, there is a third structure that is hidden in the
$Z$-function.  We are interested in describing the lightcone of a
specific point in the interior of the spacetime.  Fix a spacetime point
$x^a_0$, and consider the $S^2$ generators of its lightcone (see
(\ref{allnewcoords})), specified by:
  \begin{equation}
   \eqalign{				          
 	    u[x_0](\zeta) & =          Z(x_0^a,\zeta,\,[\sigma])  \\
       \omega[x_0](\zeta) & =     \eth Z(x_0^a,\zeta,\,[\sigma]), \\
   \bar\omega[x_0](\zeta) & = \bar\eth Z(x_0^a,\zeta,\,[\sigma]) 
	   }						  \label{allnullcones}
  \end{equation}
and the null geodesic parameter distance from $x^a_0$ to the $\zeta$
generator of ${\cal I}^+$:
  \begin{equation}					
	 R[x_0](\zeta) = \eth \bar\eth Z(x_0^a, \zeta,\,[\sigma]).
							     \label{lcgeodis}
  \end{equation}
Now, in (\ref{oldcoords}), which describes all null geodesics, at each
value of $\zeta$ we fix the null geodesic on which $x^a_0$ lies, by
substituting (\ref{allnullcones}) and (\ref{lcgeodis}) into
(\ref{oldcoords}).  This gives us, in parametric form, the full
lightcone of $x^a_0$:
  \begin{equation}					
  	x^a       [x_0](r,\zeta;\,[\sigma])=
  	x^a(	 u[x_0],
	    \omega[x_0],
  	         R[x_0]-r,\zeta;\,[\sigma]
            ),							\label{lcofx}
  \end{equation}
where the $\theta^i[x_0]$ depend on the initial data $\sigma$ and on
$\zeta$ through (\ref{allnullcones}) and (\ref{lcgeodis}). Note that
we have set $R=R[x_0]-r$, so that, at $r=0$, $x^a=x_0^a$. Thus:

\begin{quotation}
	{\em As the three parameters $(r,\zeta)$ vary, we obtain the full
	lightcone of the point $x^a_0$ in the conformal spacetime that is
	specified by $\,[\sigma]$.}
\end{quotation}

We will return to these equations and ideas later, when we 
describe the quantization of linear gravity.


\section{Vacuum Electrodynamics	}\label{vacuumaxwell}

Classical vacuum electrodynamics and classical linearized GR can be
given a parallel development in terms of characteristic data
(D'Adhamar formulation) both of which we briefly review. Though our
primary interest is in gravity, in this section, we use source-free
Maxwell theory as a model in which to introduce various ideas. As we
are only interested in giving an overview, we refrain from giving the
details \cite{wip}.

Consider Minkowski space with coordinates $x^a$ and future null
infinity ${\cal I}^+=S^2\times{\rm I\!R}$.  The characteristic data on
null infinity for the free Maxwell field is specified by the arbitrary
complex spin-wt-1 function, $A(u,\zeta)$ where, as before, $(u, \zeta)$
are the Bondi coordinates on ${\cal I}^+$.  The future lightcone of an
interior point $x^a$ intersects ${\cal I}^+$ on a (topological) sphere
$S^2(x^a)$, called the lightcone cut of $x^a$, which is described by
  \begin{equation}					 	
	u = Z_{_M}(x^a,\zeta) = x^a           \ell_a(\zeta) 
			      = x^a \eta_{ab} \ell^b,
							\label{lccut}
  \end{equation}
where $\ell^b(\zeta)$ is the null vector with components
$\frac1{\sqrt2(1+\zeta\bar\zeta)} (1+\zeta\bar\zeta, \zeta+\bar\zeta,
i(\bar\zeta-\zeta), -1+\zeta\bar\zeta)$ defining the null cone at any
point, and $Z_{_M}$ is the $Z$-function for Minkowski space.

The data $A$ restricted to the lightcone cut of $x^a$, becomes
$A_R(x^a, \zeta) = A(Z_M(x^a,\zeta), \zeta)$. The vacuum Maxwell
equations can be written in terms of a {\em real\/} scalar ``potential'',
$F(x^a, \zeta)$, which satisfies~\cite{kkn85} the equation
  \begin{equation}					
 	 \eth \bar\eth F 			= 
	 \eth \bar     A_R   (x^a, \zeta) 		+ 
     \bar\eth 	       A_R   (x^a, \zeta) 		\equiv
                       D_{_M}(x^a, \zeta)[A].		\label{maxeqn}
  \end{equation}
This is easily solved in the form
  \begin{equation}					
	 		       F(x^a, \zeta;\,[A]) 	=  
    \int_{S^2} d^2S_\eta\> G_{_M}(\zeta,\eta) 
			   D_{_M}(x^a,\eta)[A]		\label{maxfsoln}
  \end{equation}
where $G_{_M}(\zeta;\eta)$ is a known Green's function~\cite{ikn89}, and
$d^2S_\eta=-2i d\eta\wedge d\bar\eta/(1+\eta\bar\eta)^2$ is the area
form for $S^2$. $[A]$ indicates the functional dependence of the
solution on the free data. From $F$, satisfying (\ref{maxeqn}), it is
easy to construct~\cite{wip}, by differentiation, a vector potential
$\gamma_a(x^a,\,[A])$ and a Maxwell field $F_{ab}(x^a;\,[A])$ that
automatically satisfies the vacuum Maxwell equations; i.e., from $F$
we have
  \begin{equation}
  	 F         (x^a, \zeta,\,[A])       \Rightarrow 
	\gamma_a   (x^a,       \,[A])       \Rightarrow
 	 F    _{ab}(x^a,       \,[A]). 
  \end{equation}

The quantization of the Maxwell field is accomplished by constructing
operators corresponding to the data $A$ which satisfy the commutation
relations
  \begin{equation} 
	[ \widehat{     A} (u ,\zeta ), 
	  \widehat{\bar{A}\,}(u',\zeta')] = i\Delta   (u   - u'    )
					     \delta^2 (\zeta-\zeta') 
	  \widehat1,
\end{equation} 
where $\Delta(u)=sgn(u)/2$ is the skew step function, and we have
set $\hbar=c=1$. Eq. (\ref{maxfsoln}) can be used to define
\begin{equation}
     			 \widehat{F}(x^a,\zeta)    		\equiv
			           F(x^a,\zeta,[\widehat{A}\,]) =
	\int_{S^2} d^2S_\eta\> G_{_M}(    \zeta,\eta) 
			       D_{_M}(x^a,\eta) [\widehat{A}\,]
\end{equation}
and it is straightforward to find an integral representation (which can
be evaluated in closed form with considerable effort~\cite{wip}) of the
commutation relations
  \begin{equation}
		[ \widehat{F}(x ^a,\zeta ), 
		  \widehat{F}(x'^a,\zeta')
		 ]
  \end{equation}
as well as 
  \begin{equation}				\label{ccrgamma}
		[ \widehat\gamma_a(x ^a)   , 
		  \widehat\gamma_b(x'^a)
		 ] 			   \qquad\mbox{and}\qquad 
		[ \widehat{F}_{ab}(x ^a)  ,
	 	  \widehat{F}_{cd}(x'^a)
		 ].				\label{last two}
  \end{equation}
The last two (\ref{last two}) turn out to be, respectively, the
standard commutation relations of the vector potential in the Coulomb
gauge and covariant commutation relations for the $F_{ab}$. We note
that, within the formalism, the vector potential has been automatically
chosen to be in the Coulomb gauge. Alternate choices for the Green's
function $G_{_M}$ yield all other gauge choices.

We point out that the above asymptotic quantization is not merely
heuristic, but has been constructed explicitly elsewhere \cite{wip},
and a representation isomorphic to the asymptotic Fock representation
\cite{aa:aq} has been obtained, which in turn is isomorphic to the
usual Fock representation, through D'Adhamar integrals.


\section{Linearized GR: Quantization and interpretation}  \label{lingr}

The linearized Einstein equations (off Minkowski space) can be
considered as being analogous to the Maxwell equations but now for a
spin-2 field rather than for a spin-1 field.  For the moment we will
adopt this point of view and describe the linearized Einstein
equations in a fashion completely analogous to the Maxwell
description, Eqs. (\ref{maxeqn})-(\ref{ccrgamma}).  This version of
the linearized equations arises as the linearization of the
null-surface theory of GR described in Sec.\ref{vacuumaxwell}.

The data is given by a complex-valued spin-weight-2 function on ${\cal
I}^+$, namely $\sigma(u,\zeta)$ which can be given freely. The data
restricted to the lightcone cut of $x^a$ becomes $\sigma_R(x^a,
\zeta)=$~ $\sigma (Z_{_M}(x^a,\zeta),\zeta)$. The variable analogous to
the potential $F(x^a, \zeta)$ is $Z(x^a,\zeta)$ with the generalization
of (\ref{maxeqn}) to this case being~\cite{m95,nullsurface}
  \begin{equation}					
	\eth^2 \bar\eth^2 Z 				= 
	\eth^2 \bar       \sigma_R(x^a,\zeta)		+ 
    \bar\eth^2            \sigma_R(x^a,\zeta)		:= 
			       D_{_{GR}}(x^a,\zeta)[\,\sigma].   
							\label{lingreqn}
  \end{equation}
Though it is far from obvious, Eq. (\ref{lingreqn}) is completely
equivalent to the linearized vacuum Einstein equations. The solution is
easily obtained in integral form:
  \begin{equation}					
	      Z        (x^a, \zeta,[\,\sigma]) 	=  \int_{S^2}d^2\eta\>  
	      G_{_{GR}}(     \zeta,\eta) 
	      D_{_{GR}} (x^a, \eta)[\,\sigma] ,		\label{lingrsoln}
  \end{equation}
where $G_{_{GR}}$ is a known Green's function~\cite{ikn89} and the
spacetime points $x^a$ again enter as parameters in the solution
(\ref{lingrsoln}). From this, one can construct a linearized metric by
taking various derivatives of $Z$.  (See Section \ref{null-surface
theory}.)  This metric automatically satisfies the spin-2 equations ---
but in a particular gauge, namely a generalized version of the Coulomb
gauge. Analogous to the Maxwell case, we have that from $Z$ we can
calculate $h_{ab}(x^a,[\sigma])=g_{ab}-\eta_{ab}$, the deviation of the
metric from the Minkowski metric, which automatically satisfies the
linearized Einstein equations, and the linearized Weyl tensor
$C_{abcd}(x^a,[\sigma])$; i.e.,
  \begin{equation}
	Z(x^a, \zeta,[\,\sigma]) 	\Rightarrow  
	h_{ab}(x^a,  [\,\sigma])	\Rightarrow  
	C_{abcd}(x^a,[\,\sigma]).
  \end{equation}

The formal quantization of the spin-2 fields can be carried out by
representing the data as operators that satisfy the commutation
relations
  \begin{equation}						
	[\widehat     \sigma  (u, \zeta ) 	,   
	 \widehat{\bar\sigma} (u',\zeta')] 	= i
     		      \Delta  (u - u'      )
		      \delta^2(\zeta-\zeta') 
         \widehat1.					\label{ccrsigma}
  \end{equation}
From (\ref{lingrsoln}), one can define
  \begin{equation}
	\widehat{Z}       (x^a,\zeta )    \equiv 
                 Z        (x^a,\zeta, 
       [\widehat\sigma]		     ) 	=      \int_{S^2}d^2\eta\>
	         G_{_{GR}}(\zeta,\eta) 
                 D_{_{GR}}(x^a,  \eta)
       [\widehat\sigma],
  \end{equation}
and then from (\ref{ccrsigma}) one can obtain commutation relations for
the $\widehat{Z}$; i.e.,
  \begin{equation}
       [ \widehat{Z}(x ^a,\zeta ),  
         \widehat{Z}(x'^a,\zeta')
        ]
  \end{equation}
From these one can obtain the algebra of the $h_{ab}$ and $C_{abcd}$;
i.e., the commutation relations for
  \begin{equation}
	[
	 \widehat{h}_{ab}(x ^a)	,
	 \widehat{h}_{cd}(x'^a)	   
         ]            				\qquad\mbox{and}\qquad 
	[
	 \widehat{C}_{abcd}(x ^a) ,
	 \widehat{C}_{efgh}(x'^a)
	 ]			  .		
  \end{equation}
(At the present, we have an integral representation for these
commutation relations but have not evaluated the integrals in closed
form.)

Up to this point there is little difference between the Maxwell and
linearized GR theory other than the spin-1 versus spin-2. However, GR
is a theory of the geometry of spacetime. The geometric structure can
be seen in the meaning of the function, $Z(x^a,\zeta,[\,\sigma])$;
though, so far, in this section it has only played the role of a
potential for the metric (and Weyl tensor), it is an important
geometric quantity.  $Z$ is in fact the linearization of the
$Z(x^a,\zeta,[\,\sigma])$ of equation (\ref{lightcone}) of the full
theory, and as such describes the characteristic surfaces of the
linearized metric. We are now presented with a potential problem or
conundrum when trying to understand the quantization: From the point of
view of a spin-2 theory, $Z$ was a field (or potential) that could be
promoted to an operator -- but from the point of view of the geometry,
it describes a (characteristic) surface and making a ``surface'' into an operator certainly raises issues of meaning. At the very least, it
is not clear {\em a priori\/} what this may mean.

A hint as to a possible meaning of a quantum operator corresponding to
a classical definition of a surface is given by the following {\em
analogy\/}.  In ordinary quantum mechanics, consider the trajectory of
a particle.  This is described by the function $x^i(t), i=1,2,3$;
geometrically, this is a curve in spacetime. In the quantum theory,
this curve becomes a ``quantum trajectory'' $\widehat{x}^i(t)$.  That
is, we have three 1-parameter families of operators which correspond to
the position of the particle at any given value of the parameter, the
Newtonian time $t$.  The common eigenstate of the operators at time
$t$, with eigenvalues $x^i$, is the localized state
$|\,{x^i;t}\,\rangle$. For each value of the {\em classical
parameter\/} $t$, the projection of a quantum state onto the
eigenstates of $\widehat{x}^i(t)$ allows us to compute the {\em
probability\/} distribution of the position of the particle at the time
$t$. Note that $t$ is a classical parameter which specifies the
``experimental'' situation, and the probability is a density on
$\{x^i\}$, the spectrum of $\widehat{x}^i(t)$.  The interpretation of
quantum gravity that we attempt in this paper is an extension of this
very idea to the three classical geometrical entities described in Sec.~\ref{lcc}, \ref{spacetime points}, \ref{lightcones}.

\subsection{Fuzzy lightcone cuts}

Though the situation at hand is very different from, and much more
complicated than, the Newtonian particle considered above -- we are not
quantizing particle motion but geometric properties of spacetimes -- let
us first attempt a similar interpretation of $\widehat{Z}(x^a, \zeta;)
\equiv Z(x^a,\zeta;[\widehat\sigma])$. $\widehat{Z}(x^a,\zeta)$ is a
6-parameter family of operators. Like $t$, $(x^a,\zeta)$ are classical
parameters which help us define the specific ``experimental''
situation, and like the set $\{x^i\}$, the eigenvalues of $\widehat{Z}$
constitute the space of possible outcomes $\{u\}$. Classically,
(\ref{lightcone}) defined the lightcone cut of $x^a$, or the value of
$u$ at which the future lightcone from $x^a$ intersects ${\cal I}^+$ on
the generator $\zeta$.  Now, in the quantum theory, we will only have
a probabilistic interpretation. Fix a spacetime point $x^a$, a
generator of ${\cal I}^+$ labeled by $\zeta$ and a physical state
$\psi$ of ``quantum gravity''. Let $|\,{u;(x^a,\zeta)}\,\rangle$ denote
an eigenstate of $\widehat{Z}$ with eigenvalue $u$. Then we expect that
$|\langle\,{u;(x^a,\zeta)}\,|\psi\,\rangle|^2$ is the {\em
probability\/} distribution (in $u$) that the future lightcone of $x^a$
intersects ${\cal I}^+$ (at $u$) on the generator $\zeta$.

In this preliminary interpretation, it is the points, i.e. values of
$u$, along the generators of ${\cal I}^+$ which are ``fuzzy''. Thus the
lightcone cut of the point $x^a$ appears to be fuzzy.

We now develop this interpretation further. We used $Z$ to construct a
null coordinate system $(u,\omega,R)$ (in fact a sphere's worth of
them). From Sec. \ref{lcc}, let us recall the geometrical meaning of
these coordinates: $(u,\omega)$ label a null geodesic from the
generator $\zeta$ of ${\cal I}^+$, and $R$ is the geodesic distance
between ${\cal I}^+$ and a point $x^a$ on the geodesic.  In the quantum
theory, since they depend on the data through $Z$, these coordinates
will in turn be operators:
  \begin{equation}
   \eqalign{
 	\widehat{u}         (x^a,\zeta) 		& =  
			   Z(x^a,\zeta,[\widehat\sigma]),   \\   
        \widehat\omega      (x^a,\zeta) 		& = 
		      \eth Z(x^a,\zeta,[\widehat\sigma]),   \\
        \widehat{\bar\omega}(x^a,\zeta) 		& = 
                 \bar \eth Z(x^a,\zeta,[\widehat\sigma]),   \\ 
        \widehat{R}         (x^a,\zeta) 		& = 
              \eth\bar\eth Z(x^a,\zeta,[\widehat\sigma]).  \label{qoldcoords}
	   }
  \end{equation}   
Let us collectively denote them as $\widehat\theta^i$.  A number of
problems now arise.  What are the commutation relations between these
four operators?  How do the commutators depend on $\zeta$? If we fix
$\zeta$, is there a set of spacetime points such that the four geodesic
operators form a commuting set? (Note that these questions can be
answered in the linear theory~\cite{wip}.) What is the relationship
between these points?  What is clear, at such a preliminary stage, is
that a generic state of quantum gravity will not correspond to
well-defined values of $\theta^i$, since they are operators subject to
nontrivial commutation relations; in general, one can only associate
probabilities with various allowed sets of their eigenvalues. In this
sense, the lightcone cut, the angle of emittance of the null geodesic
at ${\cal I}^+$ 
and the curvature of the lightcone cut corresponding to a fixed
spacetime point are ``fuzzy''.

\subsection{Fuzzy spacetime points}

In this subsection, we construct the ``dual'' formulation (see Sec.
\ref{spacetime points}), in which the spacetime points themselves are
fuzzy.  Recall that the equations for the null coordinates of a
spacetime point $x^a$, Eq.~(\ref{allnewcoords})
  \begin{equation}
		\theta^i  = \theta^i (x^a, \zeta,[\,\sigma])
  \end{equation}
can be inverted (order-by-order in a perturbative approach, or to
first order in linear theory) to obtain the coordinates in the local chart:
  \begin{equation}					 
		x^a = x^a (\theta^i;\zeta;[\,\sigma]).  
							\label{xofq}
  \end{equation} 
In the quantization process (for the linear theory) the data, $\sigma$,
was made into operators which implied that the $\theta^i$ were all
operator functions of the $c-$numbers $(x^a, \zeta)$. Alternatively, in
the dual picture corresponding to (\ref{xofq}), we can treat the
inversion equations (\ref{xofq}) as a set of equations for the
operators $\widehat{x}^a$ as operator-valued functions of the operator
data and the $c-$numbers ($\theta^i,\zeta$):
  \begin{equation}				
	\widehat{x}^a=x^a(\theta^i;\zeta;\widehat\sigma).
							\label{qnewcoords}
  \end{equation}
From this point of view the (coordinates of the) spacetime points
themselves become operators. Let us explore the possible significance
of this.

What is the analog of a point in the quantum theory? A candidate could
be a common eigenstate $|\,{x^a;(u,\omega,R),\zeta}\,\rangle$ of the
four operators $\widehat{x}^a$; this would correspond to well-defined
values of all four coordinates, and thus a well-defined ``spacetime
point''.  Let us fix a specific null-coordinate system, by fixing
$\zeta$ corresponding to an asymptotic observer\footnote{Recall that
there many ways ($S^2$ of them) in which to reach an interior point
from ${\cal I}^+$ along null geodesics.}.  There now appear to be three
levels at which one could fail to have well-defined points in the
quantum theory:
  \begin{itemize}
     \item Consider the four coordinate operators for the same spacetime
	point; i.e., fix also $(u,\omega,R)$, and consider
	$[\widehat{x}^a,\widehat{x}^b]$. It is possible that the four
	operators $\widehat{x}^a$, corresponding to the {\em same\/}
	spacetime point, do not commute amongst themselves. Thus,
	common eigenstates $|\,{x^a;(u,\omega,R),\zeta}\,\rangle$ would
	fail to exist and, in this sense, there would be no spacetime
	points in the quantum theory. However, preliminary calculations
	suggest that they do commute.

    \item What if the coordinates of the same spacetime point {\em do\/}
	commute amongst themselves? Then, we have common eigenstates
	$|\,{x^a;(u,\omega,R),\zeta}\,\rangle$, and the existence of
	the quantum analog of a spacetime point.  However, a generic
	state of quantum gravity will not be such an eigenstate, and at
	most we will obtain a probability distribution (in $x^a$).

   \item We expect that the sets of coordinates, $\widehat{x}^a$ and
	$\widehat{x}'^a$, of two separate spacetime points will not 
	commute. Thus, even if we do find a spacetime point eigenstate,
	all the other points in the manifold are generically fuzzed
	out.
 \end{itemize}
Note that, here one obtains probability densities in $x^a$, the
eigenvalues of the spacetime point operators, whereas, in contrast to
the situation in the previous subsection, it is $(u,\omega,R)$,
together with $\zeta$ that are classical parameters which define the
``experiment'' or the measurement situation.  Thus, in this picture,
while one loses the interior of the spacetime as a distinct classical
manifold, the manifold ${\cal I}^+$ always remains free of quantum
fluctuations, and is the classical scaffolding from which quantum
measurements can be made.  What is the r\^ole of the (6 real)
parameters $(u,\omega,R,\zeta)$? In this picture, we interpret
$(u,\zeta)$ as the location of the classical, asymptotic observer; the
remaining three coordinates $(\omega,R)$ attempt to locate a specific
point on this observer's past lightcone.

\subsection{Fuzzy lightcones}

Recall from the classical theory in Sec. \ref{null-surface theory} that
we have a third interpretation (see Sec. \ref{lightcones}), one in
which the content of the Einstein equations is expressed in the
equations for the lightcones of spacetime points.  In parametric form,
the lightcone of the point $x^a_0$ is given by (\ref{lcofx}), and
through the dependence on $\sigma$, is a function on the reduced phase
space. In the quantum theory, the data, which are coordinates on the
reduced phase space, are operators. The parametric form for the
lightcone of the $c$-number $x_0$ becomes four $3-$parameter families
of operators:
\begin{equation}				\label{qlightcone}
	 \widehat{x}    ^a[x_0](r,\zeta)	=
		  x     ^a[x_0](r,\zeta;
	[\widehat\sigma]		).
\end{equation}
Again, without belaboring points we have raised before, there are
interesting issues which arise from the commutation relations between
the $\widehat{x}^a[x_0](r,\zeta)$. If we fix a spacetime point $x_0$,
are there ``lightcone eigenstates''? What physical significance can we
attribute to nontrivial commutation relations between lightcones of
different points?

Since in the above formulation the spacetime points $x_0^a$ play the
role of $c-$number classical parameters, it is natural to interpret the
uncertainties in the coordinates of the lightcones as indicating a
fuzzing out of the lightcones themselves.


\section{Discussion}					\label{conclusion}

We summarize the steps we have taken to get from the hypersurface
formulation of classical general relativity to the quantum theory in
which we have fuzzy spacetime points and other fuzzy geometric objects.
In Sec.  \ref{null-surface theory} we mentioned that the physical phase
space of general relativity can be coordinatized by the characteristic
data $\sigma$ given on ${\cal I}^+$, and comes equipped with a
symplectic structure. Einstein's equations are coded in the function
$Z$ ---a 6-parameter family of functionals on the physical phase space
$\{\sigma\}$--- which eventually determines a (conformal) Einstein
spacetime metric. $Z$ is an element of a set of four geometrical
variables on ${\cal I}^+$, namely (\ref{newcoords}). The equations for
these four geometric quantities can be inverted to yield the local
coordinates of interior spacetime points (\ref{oldcoords}), which are
also four 6-parameter families of functionals on the phase space. Next,
in Sec. \ref{lingr}, we outlined the quantum theory by promoting the
characteristic data to operators. We then constructed operators
corresponding to the four geometric variables $\theta^i$ (in section
4.1) and the spacetime points $x^a$ (in section 4.2), by promoting the
classical functions to operator valued functions of the operators
corresponding to the characteristic data. We then tried to interpret
these operators.  We discussed some consequences of the fact that a
generic physical state of ``quantum gravity'' is not a common
eigenstate of the spacetime point operators.  This discussion led us
to the conclusion that spacetime points themselves must be quantized,
and to a relatively precise formulation of the notion of a quantized or
fuzzy point.  Thus the spacetime manifold ceases to exist as a
well-defined entity in the quantum theory.

We close with some comments.
\begin{enumerate}
   \item We have an immediate {\em conceptual\/} problem: In our version of
	``quantum gravity'', though we can construct the operators
	$\widehat\theta^i$ and $\widehat{x}^a$ in linear theory (and,
	perturbatively, to higher orders), we do not yet understand how
	to do this construction (even formally) for the full
	non-perturbative theory.

   \item The quantization of the null-surface formulation can be completed
	in the case of Maxwell and in the case of linearized general
	relativity. In the later case, the Fock representation itself
	has been constructed, and work is well under way to compute all
	the required commutation relations. These calculations, which
	are lengthy but not at all conceptually difficult, are being
	carried out at present.

	In principle, at least perturbatively, the same can be done in full
	theory, the important case.  However, major technical problem
	are in the way, for examples, the factor ordering of the
	operators, their algebraic complexity, and the need to control
	the infinities.  We have not addressed these problems here.
	Rather, we have discussed a possible interpretation of the
	operators one hopes to be able to define in the full theory.

   \item The quantities $x^a(u,\omega,R,\zeta,[\sigma])$ are functionals on
	the phase space of characteristic data for the theory. Now, the
	phase space of characteristic data is essentially equivalent to
	the reduced phase space of the canonical theory. Therefore, the
	quantities $x^a(u,\omega,R,\zeta, [\sigma])$, considered as
	families of functions on the reduced phase space, are concrete
	examples of the ``evolving constants of motion'' discussed in
	Ref. \cite{evolving}.  In Ref. \cite{evolving} it was argued
	that the ``evolving constants of motion'' are the quantities
	that describe evolution in a diffeomorphism-invariant manner
	(because they are defined on the reduced phase space, and thus
	are diffeomorphism-invariant, but they code information about
	the evolution via their dependence on parameters), and
	therefore they are the quantities that must be promoted to
	operators in the quantum theory.  However, no such quantity was
	known in pure GR.  Here, we point out that ``evolving constants
	of motion'' in pure GR do in fact exist, and that they are
	realized by the quantities $x^a(u,\omega,R,\zeta, [\sigma])$.

   \item The issue of changing the coordinates by means
	of a gauge transformation is not a problem, at least
	perturbatively, since the formalism has already chosen a gauge;
	if the gauge were changed it would entail an associated change
	in the commutation relations. This is analogous to the
	situation in Maxwell theory where the vector potential can be
	quantized in the Coulomb gauge ---with specific commutation
	relations, which then are changed by a gauge transformation.

   \item The picture of quantum spacetime that emerges is still very
   vague
	and tentative.  Perhaps this view can be seen as complementary
	to the discrete quantum geometry that is emerging from the loop
	representation of quantum gravity~\cite{discreteness}, but the
	connection is certainly obscure at the moment.  We emphasize
	the fact that the picture that we have sketched is not based on
	operators which are spacetime fields (or operator valued
	spacetime distributions). In this sense, the picture is a very
	radical departure from quantum field theory: physical spacetime
	is ill-defined. On the other hand, we have kept ourselves
	rather on the conservative side as far as quantum mechanics is
	concerned, retaining the basis of its operatorial/probabilistic
	interpretation.  The reason we did this is not so much blind
	trust in quantum mechanics (which some of us do not hold), but
	rather lack of viable alternatives.  Still, we expect that the
	quantities $x^a$, derived from the null surfaces, i.e. Eq. (5),
	could represent a key to handle the mystery of a physical
	description of a quantum fluctuating geometry, free of matter.

   \item Many years ago it was pointed out by Peter Bergmann, among others,
	that since the classical theory of GR predicts the
	(pondermotive) equations of motion of its own sources, its
	quantization then should predict, or at least be closely
	associated with, the quantization of the particle orbits. Since
	the geodesic equations are limiting cases of the pondermotive
	equations, we appear to have made some contact with this idea:
	namely, there is no need to quantize the motion of particles
	separately: perhaps the quantization of our spacetime points
	already contains the quantization of classical particle
	trajectories.

   \item Notice that the approach considered here has remarkable
	similarities with some of the ideas in the twistor
	approach~\cite{twistor} to quantum gravity in both spirit and
	technique.  Twistor theory too emphasizes the quantum theoretic
	but not field theoretic viewpoint, and gives great importance
	to null surfaces and null geodesics --- as opposed to
	gravitational fields --- as descriptors of Einstein geometries.
\end{enumerate}

Work by Doplicher {\it et al}~\cite{qftqst} has recently come to our
attention, in which, motivated by semiclassical arguments, they
postulate certain commutation relations on the coordinates of
spacetime points and take preliminary steps towards a quantum field
theory on such a quantum spacetime. Connes~\cite{connes} has proposed and
developed the idea of doing quantum field theory on a noncommutative
manifold. Thus the idea of a quantum spacetime itself is not new. The
difference is that we propose to {\it derive\/} the commutation
relations on spacetime points from a quantization of (linearized) vacuum
GR, without recourse to semiclassical ideas.


\ack

The authors are deeply indebted to Abhay Ashtekar for his valuable
suggestions and criticisms.



\end{document}